\documentclass[epj]{svjour}

\usepackage{graphicx}
\usepackage{fancyhdr}
\usepackage{amsmath}
\usepackage{amssymb}
\def\makeheadbox{{
 }}
\def\mytitle{My title} 
\def\myauthors{My name}  
\def\mytype{My type of session}
\def\mysession{My session}

\def\mytitle{Constraints on the Reheating Temperature in Gravitino
  Dark Matter Scenarios} 
\def\myauthors{J.~Pradler}    
\def\mytype{Contributed Talk}    
\def\mysession{Cosmology and Astrophysics}

\newcommand{\keV}{\mathrm{keV}}

\newcommand{\GeV}{\mathrm{GeV}}
\newcommand{\TeV}{\mathrm{TeV}}
\newcommand{\seconds}{\mathrm{s}}
\newcommand{\MP}{M_{\mathrm{P}}}
\newcommand{\stau}{{\widetilde{\tau}_1}}
\newcommand{\neutralino}{{\widetilde \chi}^{0}_{1}}
\newcommand{\chargino}{{\widetilde{\chi}_1^{\pm}}}
\newcommand{\mgr}{m_{\widetilde{G}}}
\newcommand{\mstau}{m_{\widetilde{\tau}_1}}
\newcommand{\Lisix}{{}^6 \mathrm{Li}}
\newcommand{\Hefour}{{}^4 \mathrm{He}}

\newcommand{\taustau}{\tau_{\widetilde{\tau}_1}}
\newcommand{\TR}{T_{\mathrm{R}}}

\newcommand{\monetwo}{m_{1/2}}
\newcommand{\mzero}{m_{0}}
\newcommand{\tanb}{\tan{\beta}}
\newcommand{\mgut}{M_\mathrm{GUT}}
\newcommand{\Omegatp}{\Omega_{\widetilde{G}}^{\mathrm{TP}}}
\newcommand{\Omegantp}{\Omega_{\widetilde{G}}^{\mathrm{NTP}}}

\newcommand{\champ}{X^{\! -}}

\newcommand{\NLSP}{\mathrm{NLSP}}
\newcommand{\gr}{\widetilde{G}}
\newcommand{\Ydec}{Y^{\mathrm{dec}}_{\mathrm{NLSP}}}
\begin{document}

\title{CBBN in the CMSSM}
\author{Josef Pradler
\thanks{Speaker; \emph{Email:} jpradler@mppmu.mpg.de}%
 \and
 Frank Daniel Steffen
}    
%
%
\institute{Max-Planck-Institut f\"ur Physik, F\"ohringer Ring~6, D-80805 Munich, Germany}
\date{}
\abstract{%
  Catalyzed big bang nucleosynthesis (CBBN) can lead to an
  overproduction of $\Lisix$ in gravitino dark matter scenarios in which
  the lighter stau is the lightest Standard Model superpartner.  Based
  on a treatment using the state-of-the-art result for the catalyzed
  $\Lisix$ production cross section, we update the resulting constraint within
  the framework of the constrained minimal supersymmetric Standard
  Model (CMSSM).  We confront our numerical findings with recently
  derived conservative limits on the gaugino mass parameter $\monetwo$
  and the reheating temperature $\TR$.
} 


\maketitle

\section{Introduction}
\label{sec:introduction}

Big Bang nucleosynthesis (BBN) is a cornerstone of modern
cosmology that allows us to probe physics beyond the Standard Mo\-del.
It has been realized recently that the presence of massive long-lived
nega\-tive\-ly charged particles $\champ$ at the time of BBN can have
a substantial impact on the primordial light element abundances via
bound-state
formation~\cite{Pospelov:2006sc,Kohri:2006cn,Kaplinghat:2006qr,%
  Cyburt:2006uv,Hamaguchi:2007mp,Bird:2007ge,Kawasaki:2007xb,%
  Takayama:2007du,Jittoh:2007fr,Jedamzik:2007cp,Pradler:2007is}.

In scenarios in which the gravitino $\gr$ is the lightest
supersymmetric particle (LSP), a long-lived $\champ$ may be realized
if the lighter stau $\stau$ is the next-to-lightest supersymmetric
particle (NLSP). In parti\-cular, a $\stau$ NLSP can be accommodated
naturally in the framework of the constrained minimal supersymmetric
Standard Model (CMSSM) \cite{Ellis:2003dn,Cerdeno:2005eu,%
  Jedamzik:2005dh,Cyburt:2006uv,Pradler:2006hh} in which the gaugino
masses, the scalar masses, and the trilinear scalar couplings are
assumed to take on the respective universal values $\monetwo$,
$\mzero$, and $A_0$ at $\mgut\simeq 2\times 10^{16}\ \GeV$.  There
the stau emerges as the lightest Standard Model superpartner in a large part
of the CMSSM parameter space. Since the couplings of the stau to the
gravitino are suppressed by the (reduced) Planck scale, $\MP=2.4\times
10^{18}\,\GeV$, $\stau$ will typically be long-lived for conserved
$\mathrm{R}$-parity%
\footnote{For the case of broken $\mathrm{R}$-parity, see, e.g.,
  \cite{Buchmuller:2007ui,Ibarra:2007jz}.}
and thus $\stau^- = \champ$. 

Then $\stau^-$ and $\Hefour$ can form bound states,
$(\Hefour\stau^-)$, and too much $\Lisix$ can be produced via the
catalyzed (CBBN) reaction~\cite{Pospelov:2006sc}
\begin{align}
\label{eq:CBBN-reaction}
  (\Hefour\stau^-)+\mathrm{D} \rightarrow \Lisix + \stau^- \ .
\end{align}
This happens at temperatures $T \simeq 10\ \keV$ \cite{Pospelov:2006sc}
when standard BBN (SBBN) processes are frozen out. The observationally inferred bound on
primordial $\Lisix$ then severely restricts the $\stau$ abundance at
those times%
\footnote{In this work we assume a standard cosmological history in
  which $\stau$ was in thermal equilibrium before decoupling.}
and thereby the $\stau$ lifetime~$\taustau$.

For conserved R-parity, the gravitino LSP is stable and a promising
dark matter candidate.  After inflation, gravitinos are regenerated
\cite{Khlopov:1984pf} in thermal scattering of particles in the
primordial plasma.  The resulting gravitino density $\Omegatp$ will
contribute substantially to the dark matter density
$\Omega_{\mathrm{dm}}$ if the radiation-dominated epoch starts with a
high reheating temperature
$\TR$~\cite{Bolz:2000fu,Pradler:2006qh,Rychkov:2007uq}. In addition,
gravitinos are produced in stau NLSP decays with the respective
density~$\Omegantp$~\cite{Asaka:2000zh,Ellis:2003dn,Feng:2004mt}.%
\footnote{We do not include gravitino production from inflaton decays;
  cf., e.g., \cite{Endo:2007sz,Asaka:2006bv} and references therein.}

In this work we study gravitino dark matter scenarios within the
framework of the CMSSM in which $\stau$ is the NLSP. For two exemplary
parameter scans, we compute $\Omega_{\widetilde{G}} =\Omegatp +
\Omegantp$ at every point in the associated $(\mzero,\monetwo)$ planes
and compare it with $\Omega_{\mathrm{dm}}$. The $\taustau$-dependent
exclusion boundary on the stau abundance from $\Lisix$ overproduction
allows us to infer the cosmologically disfavored CMSSM region. Our
numerical findings are confronted with the recently derived
conservative limits on $\monetwo$ and $\TR$~\cite{Pradler:2007is}. We
discuss  the present status of the relevant BBN
constraints at the end of Sec.~\ref{sec:discussion}.

\section{Gravitino Dark Matter in the CMSSM}
\label{GDMintheCMSSM}

In the CMSSM, the superparticle mass spectrum is determined by
specifying $\mzero$, $\monetwo$, $A_0$, the ratio of the two MSSM
Higgs doublet vacuum expectation values, $\tanb$, and the sign of the
higgsino mass parameter $\mu$. Then either the lightest neutralino
$\neutralino$ or the lighter stau~$\stau$ with respective masses $\mstau$ and
$m_{\neutralino}$ is the lightest Standard Model superpartner and
hence the NLSP%
\footnote{A stop $\widetilde{t}_1$ NLSP is not feasible in the
  CMSSM~\cite{DiazCruz:2007fc}.} whose mass is denoted by
$m_{\NLSP}$.

With the gravitino LSP, the relic density from NLSP decays reads 
\begin{align}
  \label{eq:OmegaGntp}
  \Omegantp h^2 = \mgr \Ydec s(T_0)
  h^2/\rho_{\mathrm{c}}\ ,
\end{align}
where $\mgr$ is the gravitino mass. The quantity $\Ydec =
n^{\mathrm{dec}}_{\NLSP} / s $ denotes the NLSP yield where
$n^{\mathrm{dec}}_{\NLSP}$ is the number density at decoupling and $s
= 2\pi^2\,g_{*S}\,T^3/45$ the entropy density; $\rho_{\mathrm{c}} /
[s(T_0) h^2] = 3.6\times 10^{-9}\ \GeV$~\cite{Yao:2006px}.  We obtain
$\Ydec$ by employing the computer program \texttt{micrOMEGAs 3.17}
\cite{Belanger:2004yn} which we feed with the superparticle
mass spectrum computed with~\texttt{SuSpect 2.34}.%
\footnote{ We use the following Standard Model parameters: $m_{\mathrm{t}} =
  172.5\ \GeV$,
  $m_{\mathrm{b}}(m_{\mathrm{b}})^{\mathrm{\overline{MS}}} = 4.23\
  \GeV$,
  $\alpha_{\mathrm{s}}^{\mathrm{\overline{MS}}}(m_\mathrm{Z})=0.1172$, and
  $\alpha_{\mathrm{em}}^{-1\mathrm{\overline{MS}}}(m_{\mathrm{Z}}) =
  127.90896$.}

\begin{figure*}[t]
\begin{center}
\includegraphics[width=0.90\columnwidth]{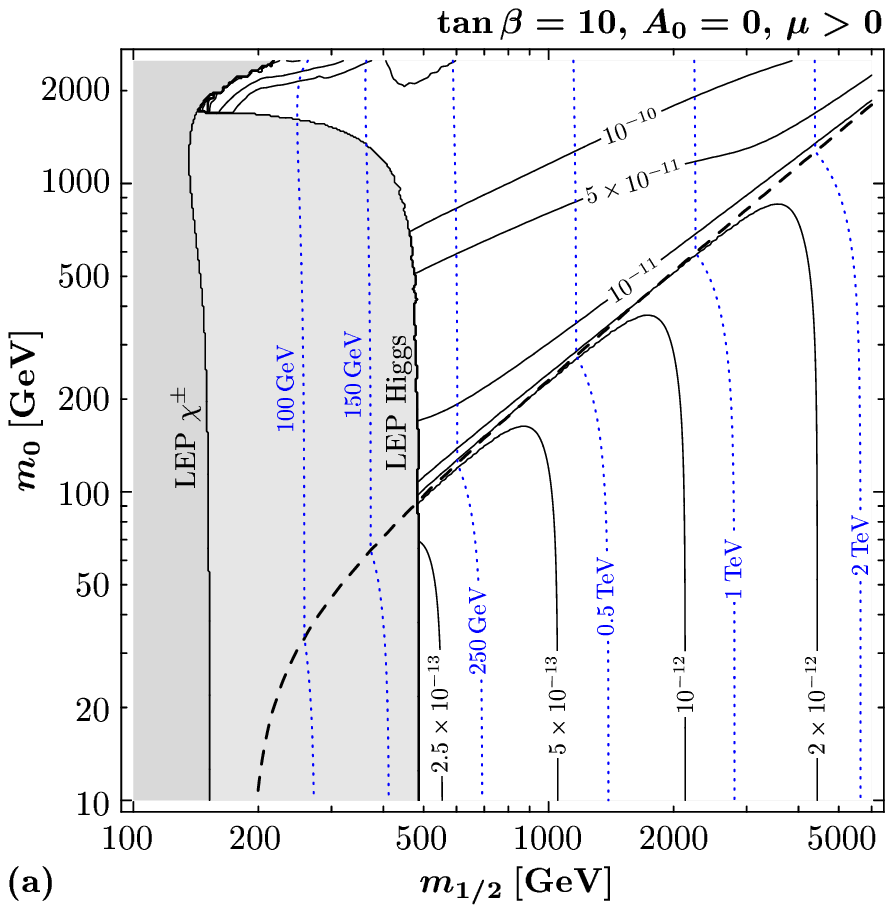} 
\hspace{0.05\columnwidth}
\includegraphics[width=0.90\columnwidth]{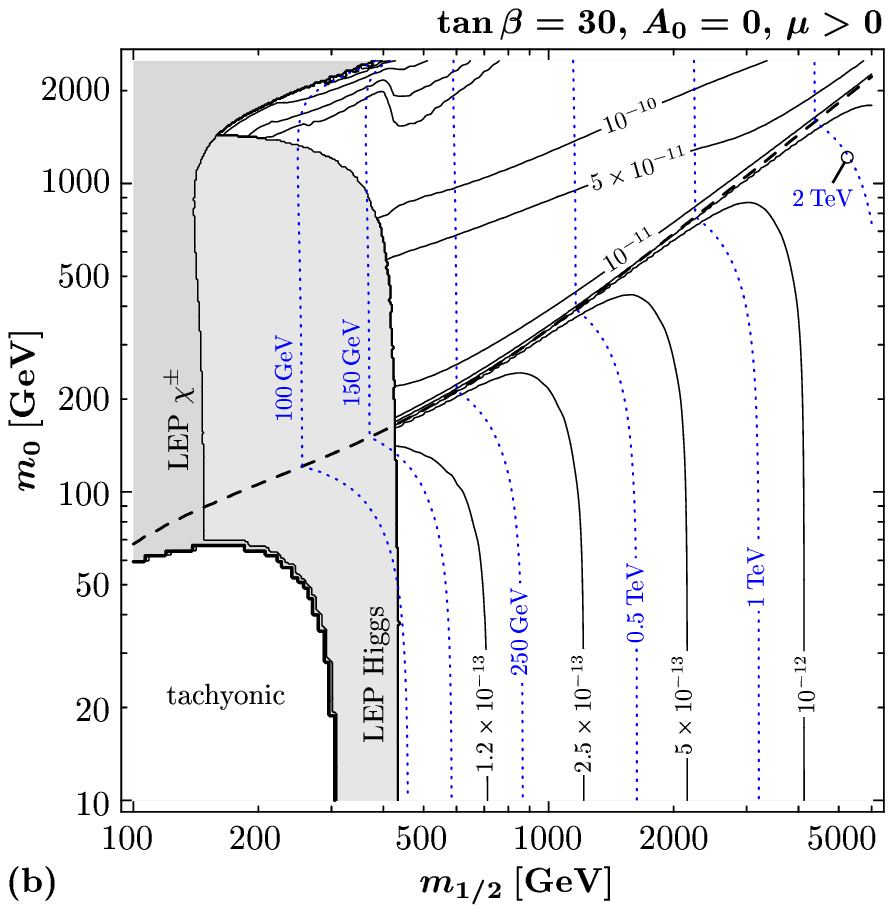} 
\vskip 0.4cm
\includegraphics[width=0.90\columnwidth]{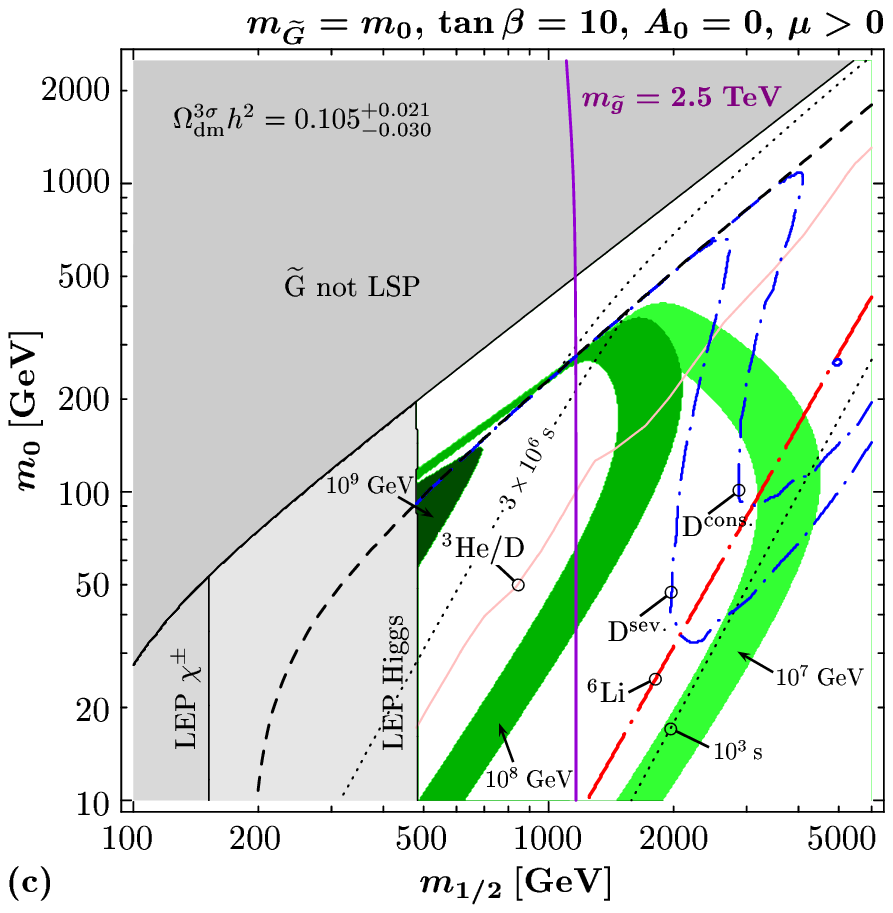} 
\hspace{0.05\columnwidth}
\includegraphics[width=0.90\columnwidth]{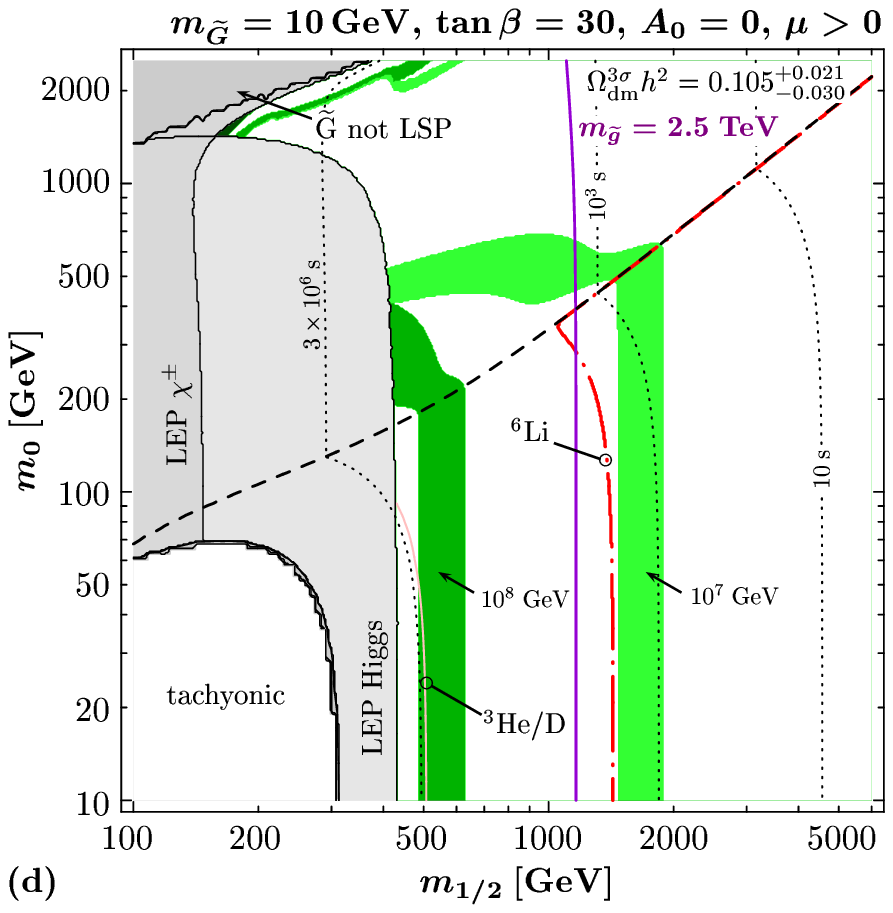} 
\caption{\small The $(\monetwo,\mzero)$ planes for $A_0=0$, $\mu>0$,
  and the choices (a,c)~$\tanb=10$ and (b,d)~$\tanb=30$.  Above
  (below) the dashed lines, $m_{\neutralino}< \mstau$
  ($\mstau<m_{\neutralino}$).  The medium gray and the light gray
  regions at small $\monetwo$ are excluded respectively by chargino
  and Higgs searches at LEP.  In the upper panels, contours of
  $Y^{\mathrm{dec}}_{\NLSP}$ (solid) and $m_{\NLSP}$ (dotted) are
  shown.  In the associated lower panels, we consider scenarios with
  (c) $\mgr=m_0$ and (d) $\mgr=10\ \GeV$. The light, medium, and dark
  shaded (green in the web version) bands indicate the regions in
  which $0.075\leq\Omega_{\widetilde{G}}h^2\leq 0.126$ for $\TR=10^7$,
  $10^8$, and $10^9~\GeV$, respectively. In the dark gray region, the
  gravitino is not the LSP.  The dotted lines show contours of the
  NLSP lifetime.  With the $\stau$ NLSP, the region to the left of the
  long-dash-dotted (red in the web version) line is disfavored by the
  primordial $\Lisix$ abundance.  Constraints from primordial D
  disfavor the $\stau$ NLSP region above the short-dash-dotted
  lines. The region to the left of the thin gray (pink in the web
  version) line is disfavored
  by~$^3\mathrm{He}/\mathrm{D}$~\cite{Kawasaki:2004qu}. On the solid
  vertical line (violet in the web version) $m_{\widetilde{g}}=2.5\
  \TeV$. }
\label{Fig:panels}
\end{center}
\end{figure*}

The upper panels in Fig.~\ref{Fig:panels} show contours of $\Ydec$
(solid) and $m_{\NLSP}$ (dotted) in the $(\monetwo,\mzero)$ plane for
$A_0=0$, $\mu>0$, (a)~$\tanb=10$ and (b)~$\tanb=30$. Above (below) the
dashed line, $m_{\neutralino}<m_{\stau}$
($m_{\stau}<m_{\neutralino}$). The medium gray and the light gray
regions at small $m_{1/2}$ are excluded respectively by the mass
bounds $m_{\chargino}>94~\GeV$ and $m_{\mathrm{h}}>114.4~\GeV$ from
chargino and Higgs searches at LEP~\cite{Yao:2006px}.  For $\tanb=30$,
tachyonic sfermions occur  at points in the
white corner labeled as ``tachyonic.''

Let us now explore the parameter space in which the relic gravitino
density matches the observed dark matter density \cite{Yao:2006px} $
\Omega_{\mathrm{dm}}^{3\sigma} h^2 = 0.105 ^ {+0.021} _{-0.030}$ where
$h$ is the Hubble constant in units of $100~\mathrm{km}\
\mathrm{Mpc}^{-1}\seconds^{-1}$. Then, $\TR$ and $\mgr$ appear in
addition to the  CMSSM parameters.  In the lower panels of
Fig.~\ref{Fig:panels}, the shaded (green in the web version) bands are
the ($\monetwo$, $\mzero$) regions in which
\begin{align}
\label{Eq:OgravitinoConstraint}
        0.075 
        \leq 
        \Omegatp h^2+\Omegantp h^2 
        \leq 
        0.126
\end{align} 
for the indicated values of $\TR$ and for (c) $\mgr =
\mzero$ and (d) $\mgr = 10\ \GeV$. For $\Omegatp$, we use
expression~(3) of Ref.~\cite{Pradler:2006qh}%
\footnote{For the definition of $\TR$, see Sec.~2 in
  Ref.~\cite{Pradler:2006hh}.}
while $\Omegantp$ is computed as explained above.  In the dark shaded
regions at larger $\mzero$, the gravitino is not the LSP.

\section{Catalyzed BBN  of \boldmath{$\Lisix$} in the CMSSM}
\label{CBBN}

We now focus on the $\stau$ NLSP region. For typical values of
$Y_{\stau^-}^{\mathrm{dec}} = \Ydec/2$
[see~Figs.~\ref{Fig:panels}~(a,b)], the amount of $\Lisix$ produced in
the CBBN reaction~(\ref{eq:CBBN-reaction}) can be as high as $\Lisix/
\mathrm{H}|_{\mathrm{CBBN}} = 10^{-7}$
\cite{Pospelov:2006sc,Hamaguchi:2007mp,Pradler:2007is}
and hence far in excess of the observationally inferred upper limit on
the primordial $\Lisix$ abundance~\cite{Cyburt:2002uv}
\begin{align}
  \label{eq:Li-obs}
  \Lisix/\mathrm{H} |_{\mathrm{obs}} \lesssim 2\times 10^{-11}.
\end{align}
For $\taustau\lesssim 5\times 10^3\ \seconds$~\cite{Pospelov:2006sc,Hamaguchi:2007mp,%
  Bird:2007ge,Takayama:2007du,Pradler:2007is}, however, the staus
decay before (\ref{eq:CBBN-reaction}) becomes too efficient so that
$\Lisix/ \mathrm{H}|_{\mathrm{CBBN}}$ can be in agreement
with~(\ref{eq:Li-obs}).
From the $\taustau$-dependent upper limit on
$Y^{\mathrm{dec}}_{\stau^-}$ computed in \cite{Pradler:2007is} by solving
the Boltzmann equations containing the state-of-the-art result
for the catalyzed $\Lisix$ production cross
section~\cite{Hamaguchi:2007mp}, we obtain the long-dash-dotted (red
in the web version) lines in the lower panels of
Fig.~\ref{Fig:panels}.  The regions to the left of theses lines are
cosmologically disfavored because of overproduction of $\Lisix$.%
\footnote{In this regard, cf.~discussion at the end of
  Sec.~\ref{sec:discussion}.}
Note that only the constraint from the primordial D abundance on
hadronic energy release \cite{Kawasaki:2004qu,Jedamzik:2006xz} in
$\stau$ decays \cite{Feng:2004mt,Cerdeno:2005eu,Steffen:2006hw} can be
more severe than the one from catalyzed $^6$Li
production~\cite{Cyburt:2006uv,Steffen:2006wx,%
  Pradler:2006hh,Kawasaki:2007xb}.%
\footnote{ Additional constraints on the primordial light elements
  from CBBN can be found in
  \cite{Cyburt:2006uv,Bird:2007ge,Kawasaki:2007xb,Jedamzik:2007cp}.}
This is shown by the short-dash-dotted (blue in the web version) lines
in panel~(c) which exclude the region in the $\stau$ NLSP region above
these lines. In panel~(d) the D constraint does not appear; for
details see \cite{Steffen:2006hw,Pradler:2006hh}. Contours of
the NLSP lifetime are shown by the dotted lines.

\section{Discussion}
\label{sec:discussion}

The position of the $\Lisix$ constraint is governed by the stau
lifetime. From the constraint $\taustau \le 5\times 10^3\ \seconds$,
conservative limits on the gaugino mass parameter,
 \begin{align}
  \label{eq:LowerLimitm12}
  \monetwo &\ge 0.9\, \TeV \left( \frac{\mgr}{ 10\ \GeV}
  \right)^{2/5} ,
\end{align}
and the reheating temperature,
\begin{align}
    \label{eq:UpperLimitTR}
  \TR &\le 4.9\times 10^7 \ \GeV \left( \frac{\mgr}{10\ \GeV}
  \right) ^{1/5} 
\end{align}
have  been derived recently~\cite{Pradler:2007is}.  Note that these
limits, obtained in the framework of the CMSSM, do only depend on
$\mgr$.

The limit (\ref{eq:LowerLimitm12}) emerges since $\mstau$ scales with
$\monetwo$ [see Figs.~\ref{Fig:panels}~(a,b)] and since $\taustau$ is
fixed once $\mgr$ and $\mstau$ are specified.  The choice $\mgr = 10 \
\GeV$ in Fig.~\ref{Fig:panels}~(d) allows for an immediate comparison
of our numerical findings with (\ref{eq:LowerLimitm12}) and
(\ref{eq:UpperLimitTR}). Only in the vicinity of the dashed line,
i.e., in the $\stau$--$\neutralino$ coannihilation region, the
position of the $\Lisix$ constraint approaches its conservative lower
limit (\ref{eq:LowerLimitm12}).  This is because $\stau$ becomes
heavier for larger $\mzero$ which shortens $\taustau$ for fixed
$\mgr$.
Contrariwise, the splitting between the actual position of the
$\Lisix$ constraint and (\ref{eq:LowerLimitm12}) is larger for smaller
$\mzero$.
At $m_0 =10\ \GeV$, this is more pronounced in
Fig.~\ref{Fig:panels}~(d) than in Fig.~\ref{Fig:panels}~(c).  This
results from the fact that the increase in $\tanb$ leads to a decrease
in $\mstau$ so that $\taustau$ becomes larger for fixed $\mgr$.

The limit (\ref{eq:UpperLimitTR}) relies on thermal gravitino
production only, $\Omegatp \sim \TR$
\cite{Bolz:2000fu,Pradler:2006qh,Rychkov:2007uq}. Thus the upper limit
on $\TR$ can only become more stringent by taking $\Omegantp$ into
account. In Fig.~\ref{Fig:panels}~(c) we have fixed $\mgr =
\mzero$. Thereby, the non-ther\-mal production (\ref{eq:OmegaGntp})
becomes more important for larger values of $\mzero$.  In addition,
$Y_{\stau}^{\mathrm{dec}}$ takes on its maximum at a given $\monetwo$
in the $\stau$--$\neutralino$ coannihilation region.
This leads to the bending of the bands~(\ref{Eq:OgravitinoConstraint})
towards lower $\monetwo$.  From Figs.~\ref{Fig:panels}~(c,d) we thus
find $\TR\lesssim 10^7\ \GeV$ confirming that~(\ref{eq:UpperLimitTR})
indeed provides a good conservative estimate. The bound $\TR \lesssim
10^7\ \GeV$ can be very restrictive for models of inflation and
baryogenesis.

Since for a $\stau$ NLSP typically $ \mzero^2 \ll \monetwo^2 $, it is
the gaugino mass parameter $\monetwo$ which sets the scale for the low
energy superparticle spectrum. Thus, depending on $\mgr$, the
bound~(\ref{eq:LowerLimitm12}) implies high values of the
superparticle masses which can be associated with a mass range that
will be difficult to probe at the Large Hadron Collider (LHC).  This
is illustrated by the vertical (violet in the web version) line in
Figs.~\ref{Fig:panels}~(c,d) which shows the gluino mass contour
$m_{\widetilde{g}} = 2.5\ \TeV$.%
\footnote{Note that the mass of the lighter stop is $
  m_{\widetilde{t}_1} \simeq 0.7 m_{\widetilde{g}} $ in the considered $\stau$ NLSP
  regions 
  with $m_{\mathrm{h}}>114.4\
  \GeV$.}

In the above considerations we have assumed a standard thermal history
of the Universe during the radiation-do\-mi\-nated epoch.  A
substantial entropy release after the decoupling of the NLSP but
before the onset of BBN may dilute
$Y^{\mathrm{dec}}_{\stau^-}$~\cite{Buchmuller:2006tt} such that
$\Lisix/\mathrm{H} |_{\mathrm{CBBN}}$ respects~(\ref{eq:Li-obs}) even
for $\taustau\gtrsim 5\times
10^3\,\seconds$~\cite{Pradler:2006hh,Hamaguchi:2007mp}. The actual
amount of entropy required, however, is model dependent.
For example, if entropy is released by an out-of-equilibrium decay of
a massive particle species $\phi$, the presence of the
energy density $\rho_{\phi}$ during NLSP decoupling can affect
$\Ydec$.  Furthermore, the branching ratio of the $\phi$ decays into
$\stau$ and/or gravitinos may be substantial; see,
e.g.,~\cite{Endo:2007sz,Asaka:2006bv}. An illustrative scenario taking
into account the former effect but neglecting the latter can be found
in \cite{Pradler:2006hh}.
As argued in \cite{Kasuya:2007cy}, however, a concrete realization of a large
entropy release by $\phi$ decays in the narrow time window after NLSP
decoupling and before BBN might be rather difficult to accomplish in
the considered scenarios.

Let us finally comment on the present status of BBN constraints on
gravitino dark matter scenarios with a long-lived charged slepton
NLSP. 
It has recently been pointed out in Ref.~\cite{Jedamzik:2007cp} that
that bound-state formation of $\champ$ with protons at $T \simeq 1\
\keV$ might well reprocess large fractions of the previously
synthesized $\Lisix$. The drop in the $\Lisix$ abundance at $t \simeq
3\times 10^6\ \seconds$ found in \cite{Jedamzik:2007cp} is rather
drastic.  As can be seen from the associated $\taustau$ contour in
Figs.~\ref{Fig:panels}~(c,d), this could reopen a $\Lisix$ conform
window in the collider friend\-ly region of low $\mstau$ and low
$m_{\widetilde{g}}$, provided that the LEP Higgs bound is respected.
Unfortunately, at the time of writing, the existing uncertainties in
the relevant CBBN nuclear reaction rates in~\cite{Jedamzik:2007cp}
make it difficult to arrive at a final conclusion on
$\Lisix/\mathrm{H}|_{\mathrm{CBBN}}$ for $\taustau \gtrsim 3\times 10^6\
\seconds $.
In this region, however, the $^3$He/D constraint on electromagnetic
energy release~\cite{Sigl:1995kk} becomes severe and can exclude
$\taustau\gtrsim 10^6~\seconds$
\cite{Cerdeno:2005eu,Cyburt:2006uv,Kawasaki:2007xb,Jedamzik:2007cp}. This
is illustrated in Figs.~\ref{Fig:panels}~(c,d) by the thin gray (pink
in the web version) line which is obtained from Fig.~42 of
Ref.~\cite{Kawasaki:2004qu} with $E_{\mathrm{vis}} = 0.3\, (\mstau^2 -
\mgr^2 + m_{\tau}^2)/2\mstau$.
Because of the strong sensitivity of the $^3$He/D constraint for
$10^6\ \seconds \lesssim \taustau \lesssim 10^7\
\seconds$~\cite{Kawasaki:2004qu,Jedamzik:2006xz}, it still remains difficult
to decide whether small cosmologically allowed islands would exist in
the CMSSM parameter space for~$\taustau > 5\times 10^3~\seconds$.%
\footnote{With a highly fine-tuned $\mstau$-$\mgr$ degeneracy leading to
  $E_{\mathrm{vis}}\to 0$, any bound on energy release can be evaded.}

\section{Conclusion}
\label{sec:conclusion}

We have considered gravitino dark matter scenarios in which $\stau$ is
the NLSP.
In exemplary CMSSM scena\-rios we have demonstrated that our recently
obtained limit~\cite{Pradler:2007is} $\TR \le 4.9 \times 10^7 \ \GeV (
\mgr/10\ \GeV ) ^{1/5}$ from catalyzed $\Lisix$ production is indeed
conservative.  In particular, taking into account $\Omegantp$, the
$\TR$ limit can become considerably more severe.  Furthermore, we have
shown explicitly that the $\Lisix$ constraint can exclude
$m_{\widetilde{g}} < 2.5\ \TeV$.  The cosmologically favored region
can thus be associated with a mass range that will be very difficult
to probe at the LHC.


\begin{thebibliography}{35}

\bibitem{Pospelov:2006sc}
M.~Pospelov, Phys. Rev. Lett. \textbf{98}, 231301 (2007)

\bibitem{Kohri:2006cn}
K.~Kohri, F.~Takayama, Phys. Rev. \textbf{D76}, 063507 (2007)

\bibitem{Kaplinghat:2006qr}
M.~Kaplinghat, A.~Rajaraman, Phys. Rev. \textbf{D74}, 103004 (2006)

\bibitem{Cyburt:2006uv}
R.H. Cyburt, J.R. Ellis, B.D. Fields, K.A. Olive, V.C. Spanos, JCAP
  \textbf{0611}, 014 (2006)

\bibitem{Hamaguchi:2007mp}
K.~Hamaguchi, T.~Hatsuda, M.~Kamimura, Y.~Kino, T.T. Yanagida, Phys. Lett.
  \textbf{B650}, 268 (2007)

\bibitem{Bird:2007ge}
C.~Bird, K.~Koopmans, M.~Pospelov (2007), \texttt{hep-ph/0703096}

\bibitem{Kawasaki:2007xb}
M.~Kawasaki, K.~Kohri, T.~Moroi, Phys. Lett. \textbf{B649}, 436 (2007)

\bibitem{Takayama:2007du}
F.~Takayama (2007), \texttt{arXiv:0704.2785 [hep-ph]}

\bibitem{Jittoh:2007fr}
T.~Jittoh et~al. (2007), \texttt{arXiv:0704.2914 [hep-ph]}

\bibitem{Jedamzik:2007cp}
K.~Jedamzik (2007), \texttt{arXiv:0707.2070v1 [astro-ph]}, \texttt{arXiv:0707.2070v2 [astro-ph]}

\bibitem{Pradler:2007is}
J.~Pradler, F.D. Steffen (2007), \texttt{arXiv:0710.2213 [hep-ph]}

\bibitem{Ellis:2003dn}
J.R. Ellis, K.A. Olive, Y.~Santoso, V.C. Spanos, Phys. Lett. \textbf{B588}, 7
  (2004)

\bibitem{Cerdeno:2005eu}
D.G. Cerdeno, K.Y. Choi, K.~Jedamzik, L.~Roszkowski, R.~Ruiz~de Austri, JCAP
  \textbf{0606}, 005 (2006)

\bibitem{Jedamzik:2005dh}
K.~Jedamzik, K.Y. Choi, L.~Roszkowski, R.~Ruiz~de Austri, JCAP \textbf{0607},
  007 (2006)

\bibitem{Pradler:2006hh}
J.~Pradler, F.D. Steffen, Phys. Lett. \textbf{B648}, 224 (2007)

\bibitem{Buchmuller:2007ui}
W.~{Buchm\"uller}, L.~Covi, K.~Hamaguchi, A.~Ibarra, T.~Yanagida, JHEP
  \textbf{03}, 037 (2007)

\bibitem{Ibarra:2007jz}
A.~Ibarra (2007), \texttt{arXiv:0710.2287 [hep-ph]}

\bibitem{Khlopov:1984pf}
M.Y. Khlopov, A.D. Linde, Phys. Lett. \textbf{B138}, 265 (1984)

\bibitem{Bolz:2000fu}
M.~Bolz, A.~Brandenburg, W.~{Buchm\"uller}, Nucl. Phys. \textbf{B606}, 518
  (2001)

\bibitem{Pradler:2006qh}
J.~Pradler, F.D. Steffen, Phys. Rev. \textbf{D75}, 023509 (2007)

\bibitem{Rychkov:2007uq}
V.S. Rychkov, A.~Strumia, Phys. Rev. \textbf{D75}, 075011 (2007)

\bibitem{Asaka:2000zh}
T.~Asaka, K.~Hamaguchi, K.~Suzuki, Phys. Lett. \textbf{B490}, 136 (2000)

\bibitem{Feng:2004mt}
J.L. Feng, S.~Su, F.~Takayama, Phys. Rev. \textbf{D70}, 075019 (2004)

\bibitem{Endo:2007sz}
M.~Endo, F.~Takahashi, T.T. Yanagida (2007), \texttt{arXiv:0706.0986 [hep-ph]}

\bibitem{Asaka:2006bv}
T.~Asaka, S.~Nakamura, M.~Yamaguchi, Phys. Rev. \textbf{D74}, 023520 (2006)

\bibitem{DiazCruz:2007fc}
J.L. Diaz-Cruz, J.R. Ellis, K.A. Olive, Y.~Santoso, JHEP \textbf{05}, 003
  (2007)

\bibitem{Yao:2006px}
W.M. Yao et~al. (Particle Data Group), J. Phys. \textbf{G33}, 1 (2006)

\bibitem{Belanger:2004yn}
G.~Belanger, F.~Boudjema, A.~Pukhov, A.~Semenov, Comput. Phys. Commun.
  \textbf{174}, 577 (2006)

\bibitem{Kawasaki:2004qu}
M.~Kawasaki, K.~Kohri, T.~Moroi, Phys. Rev. \textbf{D71}, 083502 (2005)

\bibitem{Cyburt:2002uv}
R.H. Cyburt, J.R. Ellis, B.D. Fields, K.A. Olive, Phys. Rev. \textbf{D67},
  103521 (2003)

\bibitem{Jedamzik:2006xz}
K.~Jedamzik, Phys. Rev. \textbf{D74}, 103509 (2006)

\bibitem{Steffen:2006hw}
F.D. Steffen, JCAP \textbf{0609}, 001 (2006)

\bibitem{Steffen:2006wx}
F.D. Steffen, AIP Conf. Proc. \textbf{903}, 595 (2007)

\bibitem{Buchmuller:2006tt}
W.~{Buchm\"uller}, K.~Hamaguchi, M.~Ibe, T.T. Yanagida, Phys. Lett.
  \textbf{B643}, 124 (2006)

\bibitem{Kasuya:2007cy}
S.~Kasuya, F.~Takahashi (2007), \texttt{arXiv:0709.2634 [hep-ph]}

\bibitem{Sigl:1995kk}
G.~Sigl, K.~Jedamzik, D.N. Schramm, V.S. Berezinsky, Phys. Rev. \textbf{D52},
  6682 (1995)

\end{thebibliography}
\end{document}